# A Comprehensive Performance Analysis of Proactive, Reactive and Hybrid MANETs Routing Protocols

Kavita Pandey[1], Abhishek Swaroop[2]
Comp. Sc. Deptt., JIIT, Noida

**Abstract**
A mobile Ad-hoc network (MANET) is a dynamic multi hop wireless network established by a group of nodes in which there is no central administration. Due to mobility of nodes and dynamic network topology, the routing is one of the most important challenges in ad-hoc networks. Several routing algorithms for MANETs have been proposed by the researchers which have been classified into various categories, however, the most prominent categories are proactive, reactive and hybrid. The performance comparison of routing protocols for MANETs has been presented by other researcher also, however, none of these works considers proactive, reactive and hybrid protocols together. In this paper, the performance of proactive (DSDV), reactive (DSR and AODV) and hybrid (ZRP) routing protocols has been compared. The performance differentials are analyzed on the basis of *throughput*, *average delay*, *routing overhead and number of packets dropped* with a variation of number of nodes, pause time and mobility.

*Keywords:* *MANET, proactive, reactive, hybrid.*

## 1. Introduction

Mobile Ad-hoc Networks (MANETs) are self configuring networks consisting of mobile nodes that are communicating through wireless links. There is a cooperative engagement of a collection of mobile nodes without the required intervention of any centralized access point or existing infrastructure. The nodes move arbitrarily; therefore, the network may experience unpredictable topology changes. It means that a formed network can be deformed on the fly due to mobility of nodes. Hence, it is said that an ad-hoc wireless network is self organizing and adaptive. Due to infrastructure less and self organizing nature of ad-hoc networks, it has several applications in the area of commercial sector for emergency rescue operations and disaster relief efforts. MANETs also provides a solution in the field of military battlefield to detect movement of enemies as well as for information exchange among military headquarters and so on [1]. Also, MANET provides an enhancement to cellular based mobile network infrastructure. Nowadays, it is an inexpensive alternative for data exchange among cooperative mobile nodes [2].

For communication among two nodes, one node has to check that the receiving node is with in the transmission range of source (Range of a node is defined with the assumption that mobile hosts uses wireless RF transceivers as their network interface), if yes, then they can communicate directly otherwise, with the help of intermediate nodes communication will take place. Each node will act as a host as well as a router. All the nodes should be cooperative so that exchange of information would be successful. This cooperation process is called as routing [3, 4].

Due to the presence of mobility, the routing information will have to be changed to reflect changes in link connectivity. There are several possible paths from source to destination. The routing protocols find a route from source to destination and deliver the packet to correct destination. The performance of MANETs is related to efficiency of the MANETs routing protocols [5] and the efficiency depends on several factors like convergence time after topology changes, bandwidth overhead to enable proper routing, power consumption and capability to handle error rates.

The figure 1 shows the prominent way of classifying MANETs routing protocols. The protocols may be categorized into two types, Proactive and Reactive. Other category of MANET routing protocols which is a combination of both proactive and reactive is referred as Hybrid.

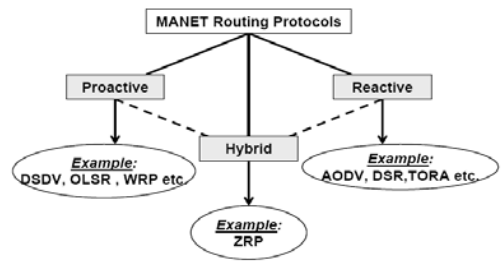

Figure 1 Classification of MANET routing protocols

**Proactive routing protocols:** In it, all the nodes continuously search for routing information with in a network, so that when a route is needed, the route is already known. If any node wants to send any information to another node, path is known, therefore, latency is low. However, when there is a lot of node movement then the cost of maintaining all topology information is very high [6].

**Reactive Routing protocols:** Whenever there is a need of a path from any source to destination then a type of query reply dialog does the work [7, 8].Therefore, the latency is high; however, no unnecessary control messages are required.

**Hybrid routing protocols:** These protocols incorporates the merits of proactive as well as reactive routing protocols. A hybrid routing protocol should use a mixture of both proactive and reactive approaches. Hence, in the recent years, several hybrid routing protocols are proposed like ZRP, ZHLS, SHARP and NAMP etc [7, 9].

In recent years, a variety of routing protocols have been proposed and a comparative analysis of routing protocols has been done either on the basis of simulation results by different simulators like OPNET, NS2, OMNET++ etc. or analytically. In some cases, the comparative analysis is done between reactive routing protocols based on some performance metrics and in other cases between proactive routing protocols. Few researchers have done the simulation based comparison between on demand and table driven routing protocols. The present paper comparatively analyzes all three categories of MANETs routing protocols namely, proactive, reactive and hybrid protocols. In order to compare the protocols, we selected the representative protocols from each category; DSDV from proactive, ZRP from hybrid, and AODV and DSR from the reactive. The performance metrics considered are *throughput*, *average delay*, *routing overhead and number of packets dropped*. The performance differentials are analyzed by varying number of nodes, pause time and mobility using NS2 simulator.

The rest of the paper is organized as follows. The related work has been discussed in section 2. Section 3 provides a brief summary about these protocols. In the section 4, the simulation environment, performance metrics used and results have been discussed. Section 5 concludes the present exposition.

## 2. Related Work

While most of the work done related to the performance comparison of MANETs routing protocols includes either purely reactive protocols or purely proactive protocols. Some researchers have done a comparative study on reactive and proactive or reactive and hybrid protocols. The table 1 summarizes the work done by various researchers related to performance analysis of MANETs routing protocols.

Table 1: Related work

| Author Name Reference | Protocols Used | Simulator | Performance Metrics | Variable Parameters |
|---|---|---|---|---|
| Guntupalli et al. [5] | DSDV, DSR, AODV | NS2 | Average End to End Delay, Normalized Routing Load, Packet Delivery Ratio | Number of Nodes, Speed, Pause time, Transmission Power |
| Yogesh et al. [10] | AODV, DSR | GLOMOSIM | Packet Delivery Ratio, End to End Delay, Normalized routing overhead | Number of nodes, Speed, Pause time |
| Chenna et al [11] | DSDV, AODV, DSR, TORA | NS2 | Throughput, Routing Overhead, Path Optimality, Packet Loss, Average delay | Traffic Loads, Movement patterns |
| G. Jayakumar et al. [3] | AODV, DSR | NS2 | Packet Delivery Ratio, Normalized Routing Load, MAC load and Average End to End Delay | Number of Sources, Speed, Pause time |
| Birdar et al.[2] | AODV, DSR | NS2 | Packet Delivery Ratio, Routing Overhead, Normalized Routing Overhead and Average End to End Delay | Speed |
| Kapang et al. [1] | AODV, DSR, DSDV | NS2 | Packet Delivery Ratio, Average End to End Delay and Routing Overhead | Pause Time |
| Vijayalaskhmi et al. [12] | DSDV, AODV | NS2 | Packet Delivery Fraction, Average End to End Delay and Throughput | Number of Nodes, Speed, Time |
| Shaily et al. [13] | AODV, DSR, ZRP | QualNet | TTL based Hop Count, Packet Delivery Ratio and Average End to End Delay | Pause Time |
| Li Layuan et al. [14] | DSDV, AODV, DSR, TORA | NS2 | Average Delay, Jitter, Routing Load, Loss Ratio, Throughput and Connectivity | Network Size |

It is evident from table 1 that, no one has presented the comparison of performance differentials among proactive, reactive and hybrid protocols.

## 3. MANETs Routing Protocols

### 3.1 DSDV (Destination Sequence Distance Vector)

It is a proactive routing protocol and based on the distributed Bellman-Ford Algorithm. The improvement from distance vector in wired routing protocol is in the terms of avoidance of routing loops. Each node maintains a routing table which has the list of all the possible destinations and number of routing hops to reach the destination. Whenever some packet comes to node, routing table is to be consulted to find the path. DSDV uses a concept of sequence numbers to distinguish stale routes from new routes and the sequence number is generated by the destination node. To maintain consistency in routing table, DSDV sends routing updates periodically [15]. Therefore, a lot of control message traffic which results in an inefficient utilization of network resources. To overcome this problem, DSDV uses two types of route update packets: *full dump, incremental packets* [16, 17].

3.2 DSR (Dynamic Source Routing)

DSR is a pure reactive routing protocol which is based on the concept of source routing. DSR protocol is composed of two important phases: *route discovery* and *route maintenance.* DSR does not employ any periodic routing advertisement packets, link status sensing or neighbor detection packets [15]. Therefore, the routing packet overhead is less because of its on-demand nature. Every node maintains a route cache to store recently discovered paths. Whenever a route is required for a particular destination then that particular node will consult route cache to determine whether it has already a route to the destination or not. If available route is not expired then that route will be used otherwise a route discovery process is initiated by broadcasting the *route request packet (RREQ)*. When any of the nodes receives RREQ packet, the node will check from their cache or from their neighbors whether it knows a route to the destination. If it does not, the node will add its own address to the route record of the packet and forwards it to their neighbors. Otherwise; a *route reply packet (RREP)* is generated that is unicast back to the original source.

Due to dynamic nature of the environment, any route can fail anytime. Therefore, the route maintenance process will constantly monitors the network and notify the other nodes with the help of route error packets as well as route cache would be updated [16, 18].

3.3 AODV (Ad-hoc On-demand Distance Vector)

AODV algorithm is pure reactive in nature and it contains the properties of both DSR and DSDV protocols. AODV algorithm is an improvement on DSDV in the sense that it minimizes the number of broadcasts. AODV borrows the concept of hop by hop routing, sequence numbers, periodic beacon messages from DSDV protocol [15]. Like DSR, it is on-demand protocol but unlike source routing. When a node wants to send a message to destination node, first it will check whether it has a valid route to the destination or not. If not, then it broadcast a *route request packet* (RREQ) to its neighbors which then forwards the request to their neighbors and so-on, until either it reaches to the intermediate node which has a valid route for the destination or the destination node. AODV uses destination sequence numbers to ensure that it contains most recent information and all routes are loop-free. Once the route request has reached the destination or an intermediate node with a valid route, the destination/intermediate node responds by unicasting a *route reply (RREP) message* back to the neighbor node from which it first received the RREQ [16, 19]. The route maintenance process in AODV is performed with the *route error (RERR) message*. *Hello messages* are used for periodic local broadcast to maintain the local connectivity of the network.

3.4 ZRP (Zone Routing Protocol)

Zone routing protocol is a hybrid protocol. It combines the advantages of both proactive and reactive routing protocols. A routing zone is defined for every node. Each node specifies a zone radius in terms of hops. Zones can be overlapped and size of a zone affects the network performance. The large routing zones are appropriate in situations where route demand is high and / or the network consists of many slowly moving nodes [15]. On the other hand, the smaller routing zones are preferred where demand for routes is less and /or the network consists of a small number of nodes that move fast relative to one another. Proactive routing protocol works with in the zone whereas; reactive routing protocol works between the zones.

ZRP consists of three components: **1) the proactive Intra zone routing protocol (IARP), 2) the reactive Inter zone routing protoc ol (IERP) and 3) Bordercast resolution protocol (BRP).** Each component works independently of the other and they may use different technologies in order to maximize efficiency in their particular area. The main role of IARP is to ensure that every node with in the zone has a consistent updated routing table that has the information of route to all the destination nodes with in the network. The work of IERP gets started when destination is not available with in the zone. It relies on bordercast resolution protocol in the sense that border nodes will perform on-demand routing to search for routing information to nodes residing outside the source node zone [20].

4. Simulation

There are several simulators available like OMNET++, QualNet, OPNET and NS2. Here, NS2 is used for simulation experiments since it is preferred by the networking research community. NS2 is an object oriented simulator, written in C++ and OTcl (Object oriented Tool command language) as the frontend. If the components have to be developed then both Tcl (Tool command language – scripting language) and C++ have to be used. In this section, we have described about the performance metrics and implementation details of all four

MANETs routing protocols namely, DSDV, DSR, AODV and ZRP.

4.1 Performance Metrics

The following performance metrics are considered for evaluation of MANETs routing protocols:
**Throughput:** the ratio of data packets received to the destination to those generated by source.
**Average Delay:** it includes all possible delays caused by buffering during route discovery latency, queuing at the interface queue, retransmission delays at the MAC, and propagation and transfer times. It is the average amount of time taken by a packet to go from source to destination. [19]
**No. of packets dropped:** it is the number of packets lost by routers at the network layer due to the capacity of buffer or the packet buffering time exceeds the time limit.
**Routing Overhead:** it is the number of routing packets which would be sent for route discovery and maintenance. All the above mentioned performance metrics are quantitatively measured. For a good routing protocol, throughput should be high where as other three parameters value should be less.

4.2 Implementation

The simulation parameters considered for the performance comparison of MANETs routing protocols are given below:

Table 2: Simulation Setup parameters

| Platform | Linux, Fedora core 9 |
|---|---|
| NS Version | ns-allinone-2.34 |
| Protocol | DSDV, AODV, DSR , ZRP |
| Mobility Model | Random way Point |
| Area | 500 * 500 m |
| Experiment Duration | 150 sec |
| Traffic Type | CBR |
| Radio Propagation | TwoRayGround |
| MAC layer Protocol | Mac/802_11 |
| Packet size | 512 bytes |
| Antenna type | Antenna/OmniAntenna |
| Number of nodes | 10, 20, 30, 40, 50 |
| Maximum Speed | 5, 10, 15, 20, 25 m/s |
| Pause time | 10, 50, 100, 150, 200 |

NS2 provides the implementation of DSDV, AODV and DSR protocols. However, for ZRP, a patch has been integrated into NS2 package [21, 22]. The Tcl code has been written to set up the network components which includes the parameters defined in Table 2. For **traffic model**, cbrgen utility has been used which creates CBR and TCP traffic connections between nodes [23]. The different traffic files have been generated by varying the number of nodes with CBR traffic source at a rate of 4 packets / sec and keeping maximum number of connections as 20 to 40.

For **mobility model**, setdest utility has been used to create node positions and their movements [23]. In order to perform simulation experiments, twenty five different scenario files have been generated by varying the number of nodes and pause time and keeping other values constant. Other twenty five scenario files have been generated by varying the number of nodes and maximum speed by keeping the pause value as 2 seconds. Pause time, Max speed and number of nodes are varied according to the table 2.

The **Tcl** file generates different trace files according to different MANETs routing protocols. In order to test the behavior of different protocols, the trace files have been parsed with the help of programs written in **Python** language to extract the information needed to measure the performance metrics. After getting the values of different performance metrics according to different routing protocols, **XGraph** utility is used to plot the graphs. **Network Animator (NAM)** is used to graphically visualize the simulation [24, 25,26].

4.3 Results

Simulation Results have been presented in the group of four figures where each figure is corresponding to one performance metric. The performance metrics considered are *throughput, average delay, number of packets dropped* and *routing overhead*. In all graphs, x-axis specifies the number of nodes and y-axis indicates the value of performance parameter.

We have presented the analysis of results according to different performance metrics. With each performance metric, results are analyzed by changing the number of nodes, speed and pause time. The group of graphs numbered from 1 to 5 shows the simulation results by varying the pause values and number of nodes, however, the maximum speed is kept as constant that is 2 m/s. Whereas, the next group of graphs numbered from 6 to 10 shows the simulation results by keeping the pause value constant i.e., 2 seconds and varying the maximum speed and number of nodes. The minimum speed is taken as 1 m/s, so, that nodes will move with an average speed.

**Throughput:** It is evident from the results that throughput of AODV is better as compared to other protocols. Moreover, the change in the pause value does not have any effect on AODV performance. Generally, for all the protocols, by increasing the number of nodes, throughput also increases. In DSDV, initially (before the convergence of roots), some packets are sent and get dropped, therefore, it has low throughput as compared to AODV and DSR. With pause value 200 and number of nodes 10, the throughput of DSDV is zero. The throughput of ZRP does not change even on changing the pause value or speed or the number of nodes. The reason behind this phenomenon may be the fixed zone radius. On changing the pause value, the throughput of DSR has an oscillating behavior. One possible reason is that DSR uses route cache and the routes stored in the cache might become stale after sometime. However, by increasing the speed, the throughput of DSR decreases. This can be due to the mobility of nodes which may increase the chances of path failure.

**Average Delay:** From the graphs related to average delay, it can be seen that AODV and ZRP has higher average delay where as other two protocols experiences less delay. In DSR, due to the caching of routes, the average delay is reduced. However, as the number of nodes increases, DSR exhibits significantly higher delay then other three protocols. This may be due to the increasing node density because of which the number of data sessions increases which leads to increased end to end delay. Average delay of DSDV is less in comparison to other three protocols since it is a proactive protocol. The routes for all the destinations are maintained in routing tables. When speed increases, there is no effect on average delay. However, as the number of nodes increases, the delay increases due to time consumed in computation of routes, however, once routes have been created; the delay becomes less as evident from the graphs. Since, AODV is a reactive protocol, the routes are created on demand; therefore, it experiences a higher delay. As speed increases, spikes in AODV are higher; however, for higher mobility, AODV has less delay as compared to previous values. In case of ZRP, initially, when number of nodes is less, it has a higher delay because of the route creation and table maintenance, then the delay decreases and after that it gives a mediocre performance which is expected because of its hybrid nature. In ZRP, on increasing the speed and the number of nodes, the delay increases because of difficulty in setting routes due to contention and high mobility.

**Number of Dropped Packets:** In DSDV protocol, more number of packets gets dropped as compared to other protocols. Generally, the value is higher when the number of nodes is less. The reason may be sending the data packets before convergence of routes. DSR and AODV experience a similar behavior that dropped packets are less, which specifies their high reliability. However, in DSR the number of dropped packets is marginally less in comparison to AODV. This reduction may be due to the fact that DSR maintains route cache. Generally, by increasing the pause value, number of dropped packets also increases. Initially, ZRP has less number of dropped packets; however, as number of nodes increases, there is a sharp increase in the value. In every protocol, the number of dropped packets increases on increasing the speed due to difficulty in path creation. With an increase in speed, descending order of performance corresponding to number of dropped packets is ZRP, DSR, AODV, and DSDV. On increasing the speed, the number of dropped packets in DSDV protocol is high since it is a table-driven routing protocol.

**Routing Overhead:** ZRP and AODV have more routing overhead in comparison to DSR and DSDV routing protocols. In DSR, the routes are maintained only between the nodes those want to communicate as well as a single route discovery may yield many routes to the destination, therefore, the routing overhead is less. Where as, in DSDV, the concept of table maintenance reduces the routing overhead. In ZRP, maintaining the zone radius as well as electing the border nodes and switching from proactive to reactive or vice-versa, more number of control packets are needed. As number of nodes increases, the routing overhead increases because of increasing node density. In ZRP and AODV, routing overhead increases by a large amount where as, in DSDV and DSR, it increases marginally. In DSDV and DSR, there is not a significant effect of pause value or speed value. Where as, in ZRP overhead is reduced by a small amount with respect to increase in pause value and speed. This scenario is reversed in AODV protocol, that is by increasing the speed and pause value, overhead also increases. At the last, it can be concluded that the routing overhead increases with increasing number of nodes; however, a change in pause value or speed does not adversely affect the performance.

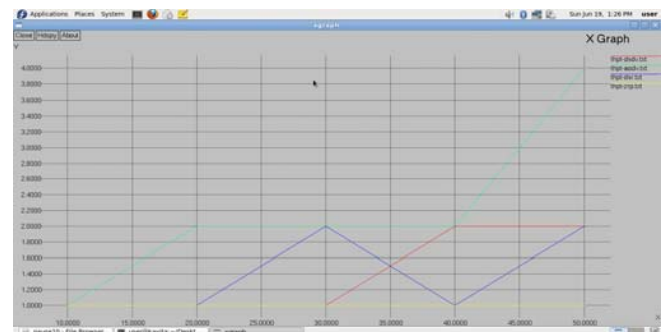
Graph 1.1- Throughput, pause 10 and varying number of nodes

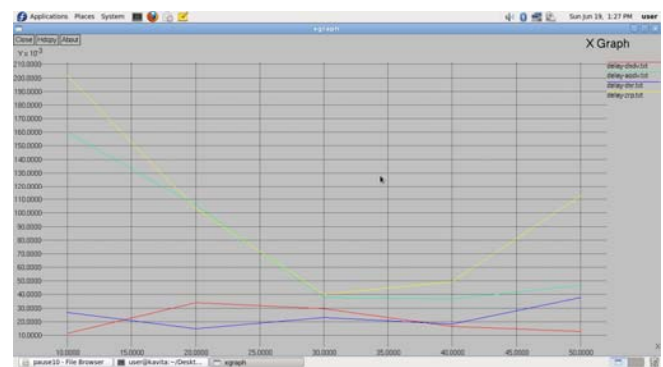
Graph 1.2- Average Delay, pause 10 and varying number of nodes

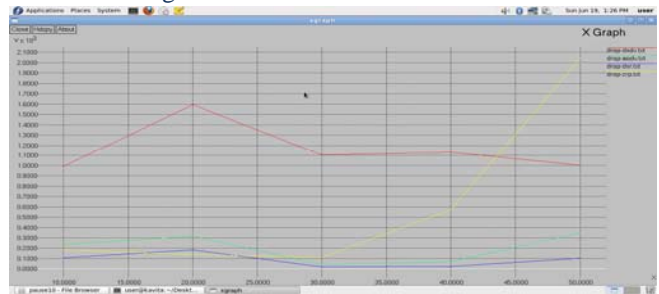
Graph 1.3- Dropped Packets, pause 10 and varying number of nodes

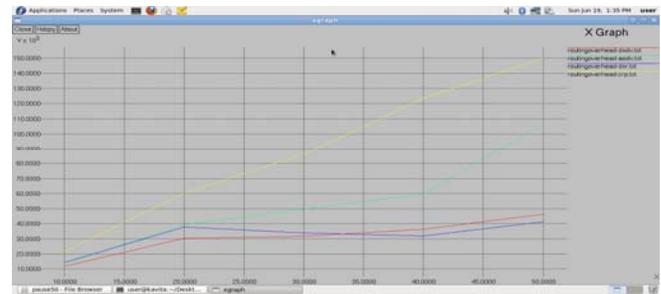
Graph 2.4- routing Overhead, pause 50 and varying number of nodes

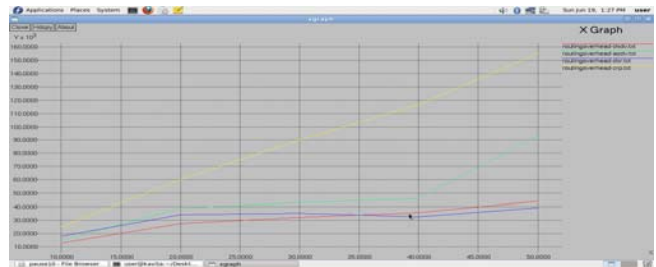
Graph 1.4- Routing Overhead, pause 10 and varying number of nodes

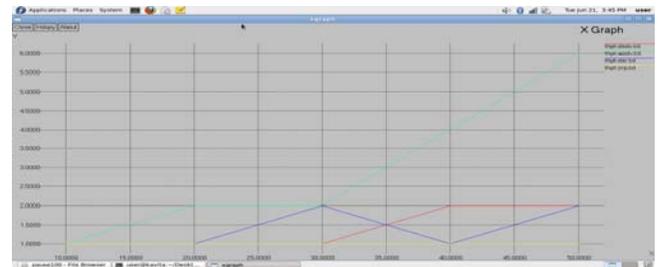
Graph 3.1- Throughput, pause 100 and varying number of nodes

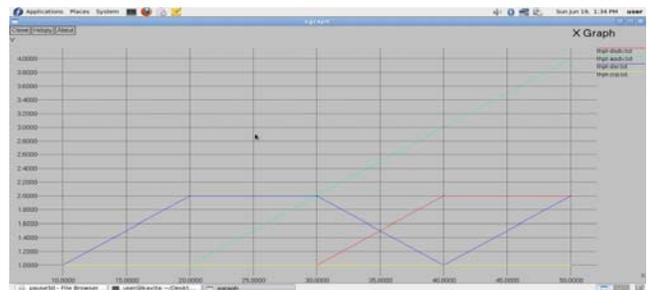
Graph 2.1- Throughput, pause 50 and varying number of nodes

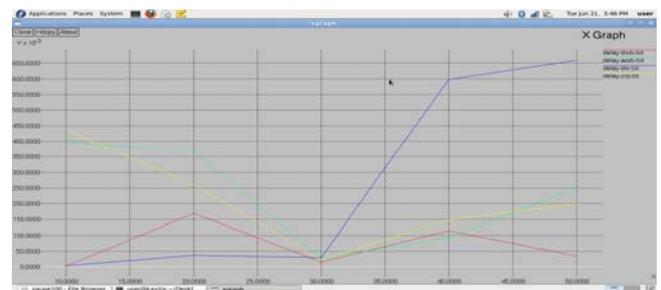
Graph 3.2- Average Delay, pause 100 and varying number of nodes

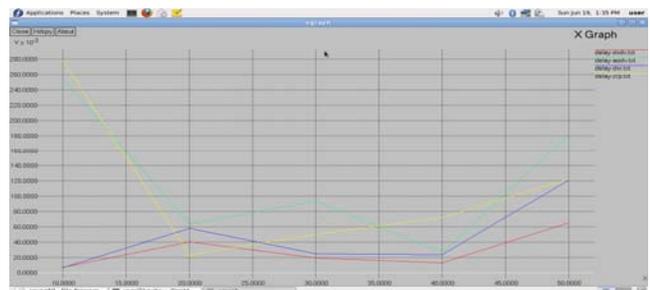
Graph 2.2- Average Delay, pause 50 and varying number of nodes

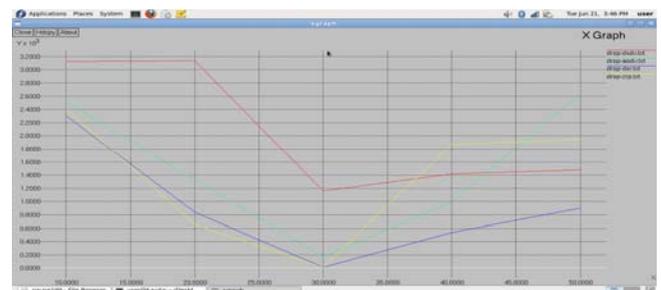
Graph 3.3- Dropped packets, pause 100 and varying number of nodes

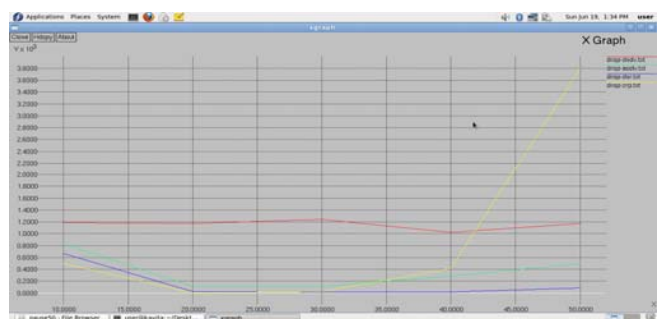
Graph 2.3- Dropped Packets, pause 50 and varying number of nodes

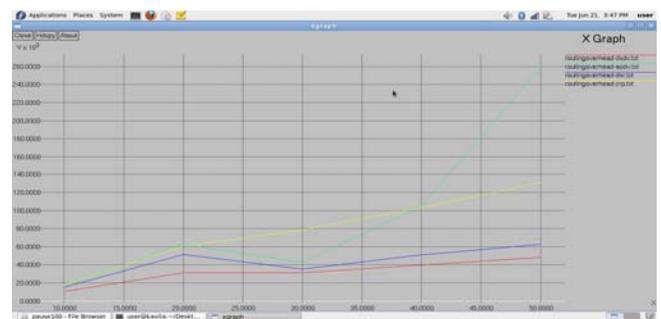
Graph 3.4- Routing Overhead, pause 100 and varying number of nodes

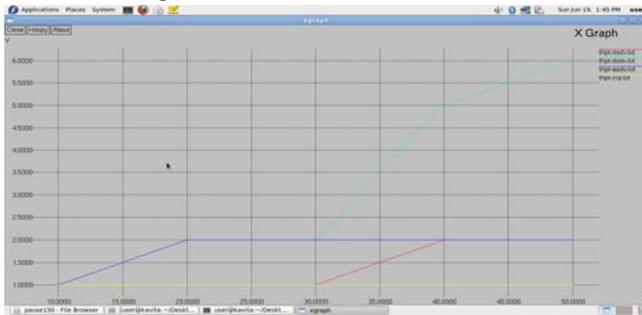
Graph 4.1: Throughput, pause 150 and varying no of nodes

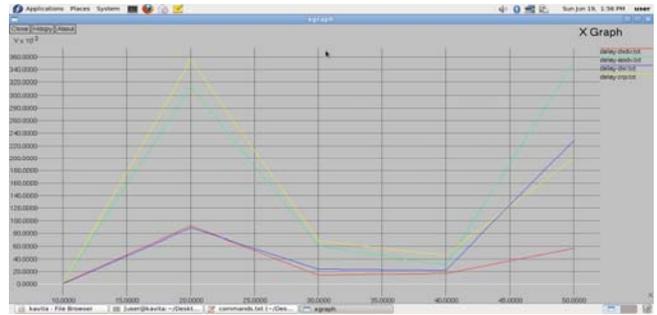
Graph 5.2: Average Delay: pause 200 and varying no of nodes

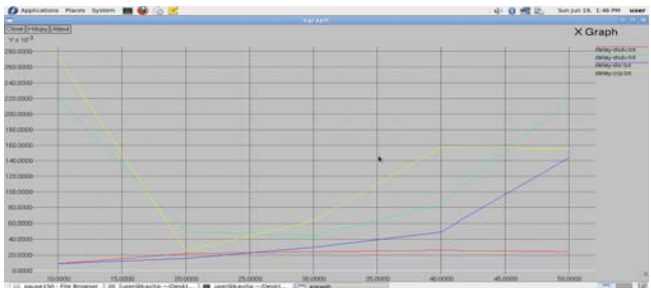
Graph 4.2: Average Delay, pause 150 and varying no of nodes

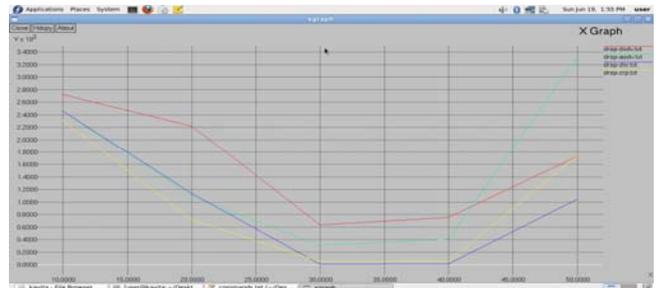
Graph 5.3: Dropped Packets: pause 200 and varying no of nodes

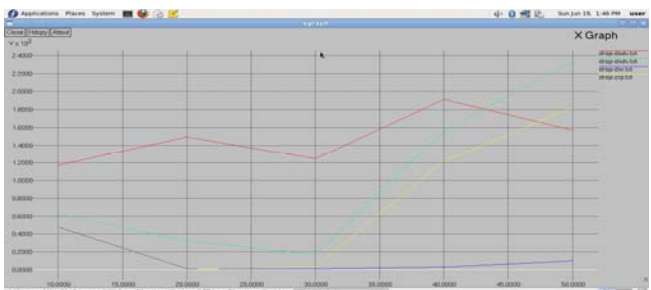
Graph 4.3: Dropped Packets, pause 150 and varying no of nodes

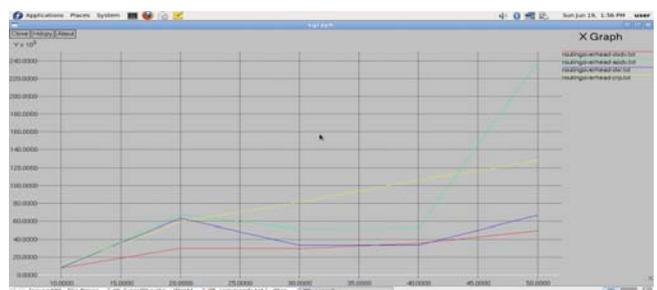
Graph 5.4: Routing Overhead: pause 200 and varying no of nodes

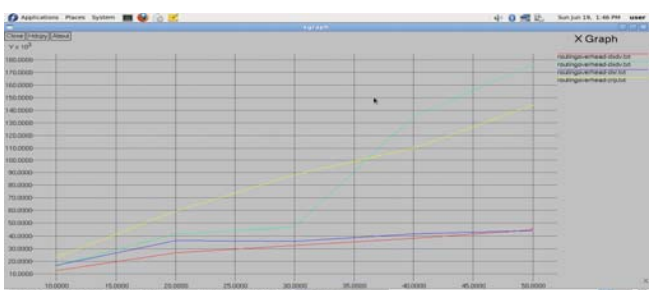
Graph 4.4: Routing Overhead, pause 150 and varying no of nodes

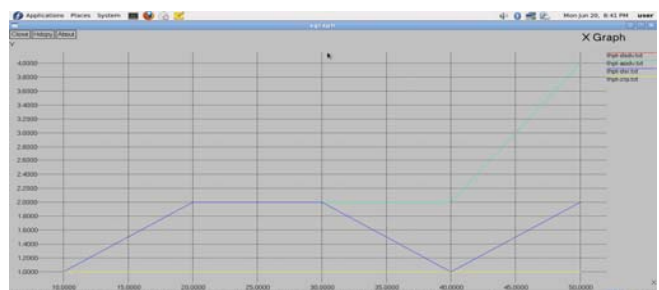
Graph 6.1: Throughput, speed 5 and varying number of nodes

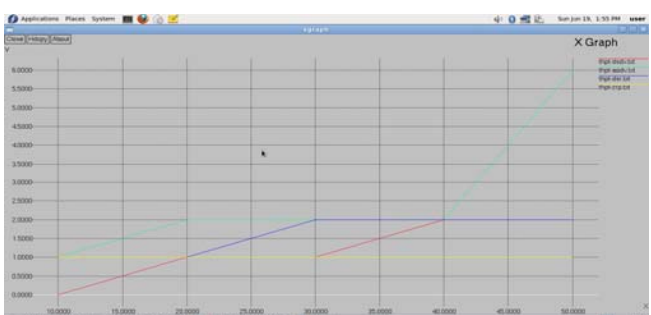
Graph 5.1: Throughput: pause 200 and varying no of nodes

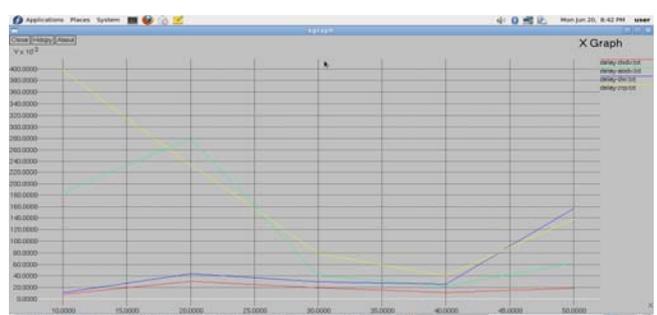
Graph 6.2: Average Delay, speed 5 and varying number of nodes

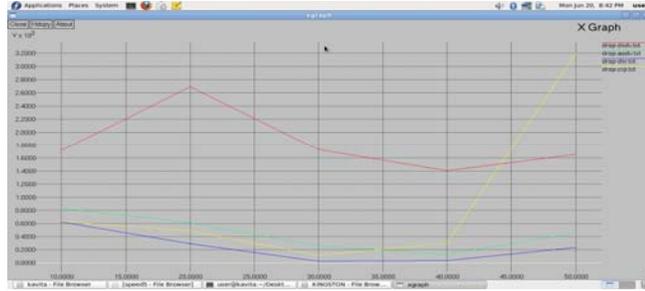
Graph 6.3: Dropped Packets, speed 5 and varying number of nodes

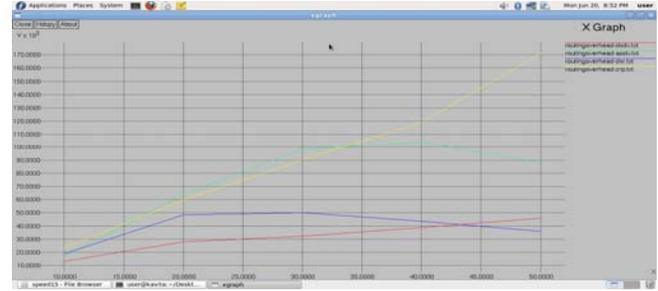
Graph 7.4: Routing Overhead, speed 10 and varying number of nodes

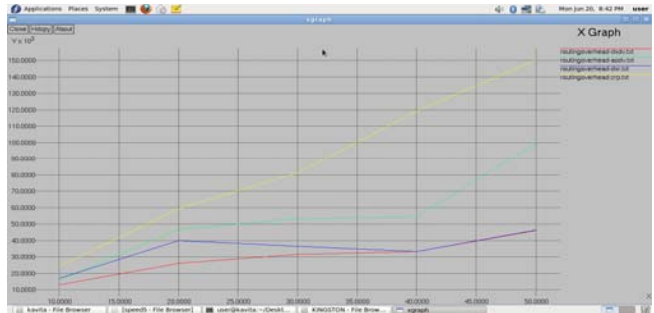
Graph 6.4: Routing Overhead, speed 5 and varying number of nodes

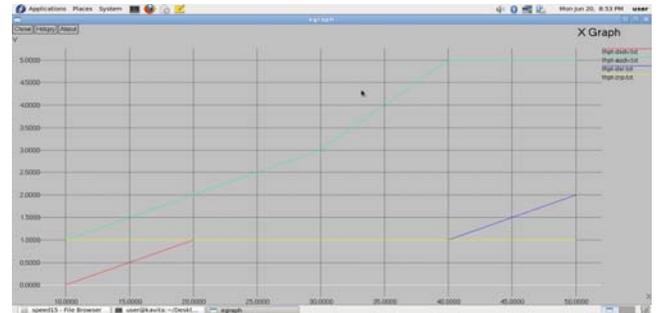
Graph 8.1: Throughput, Speed 15 and varying number of nodes

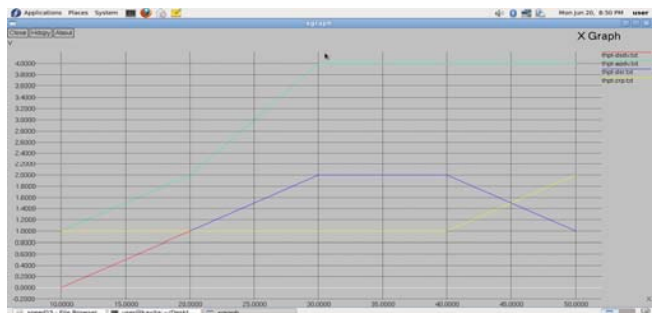
Graph 7.1: Throughput, speed 10 and varying number of nodes

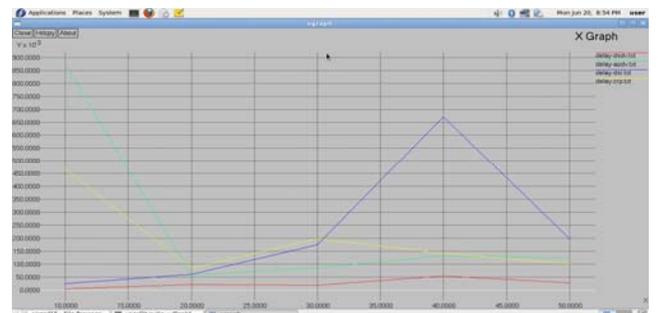
Graph 8.2: Average Delay, Speed 15 and varying number of nodes

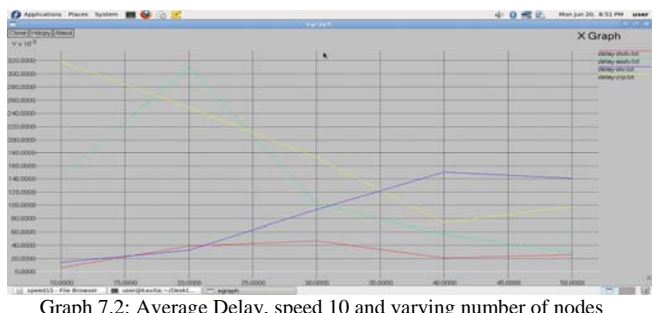
Graph 7.2: Average Delay, speed 10 and varying number of nodes

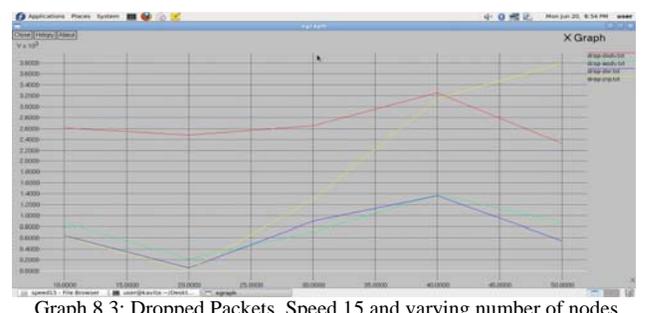
Graph 8.3: Dropped Packets, Speed 15 and varying number of nodes

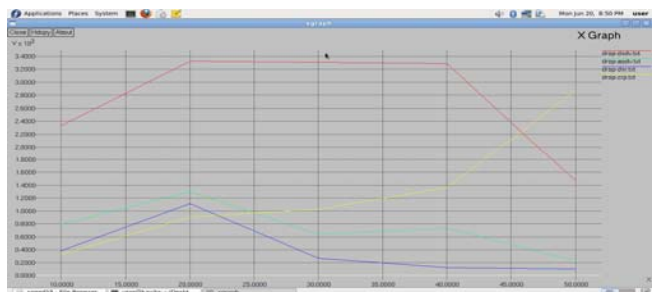
Graph 7.3: Dropped Packets, speed 10 and varying number of nodes

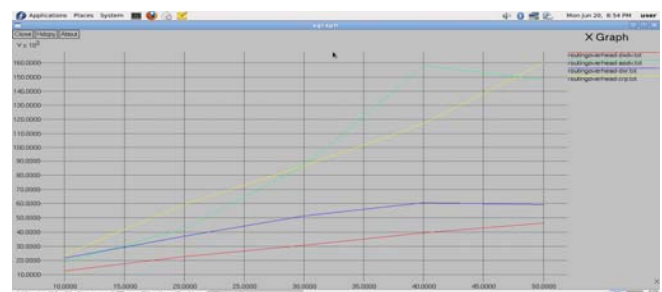
Graph 8.4: Routing Overhead, Speed 15 and varying number of nodes

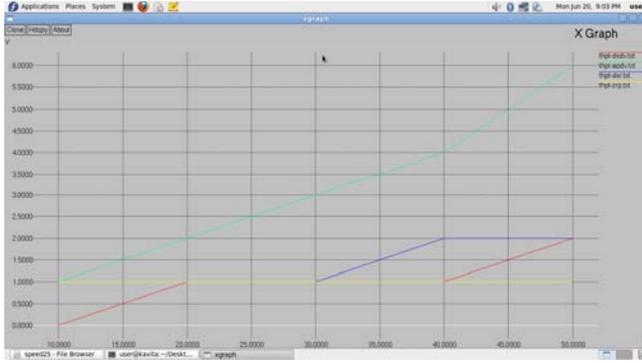
Graph 9.1: Throughput, Speed 20 and varying number of nodes

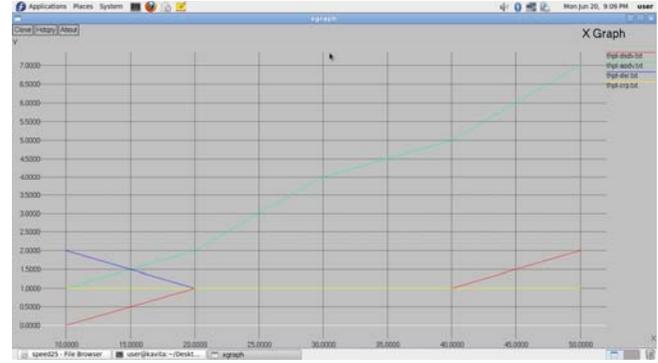
Graph 10.1: Throughput, Speed 25 and varying number of nodes

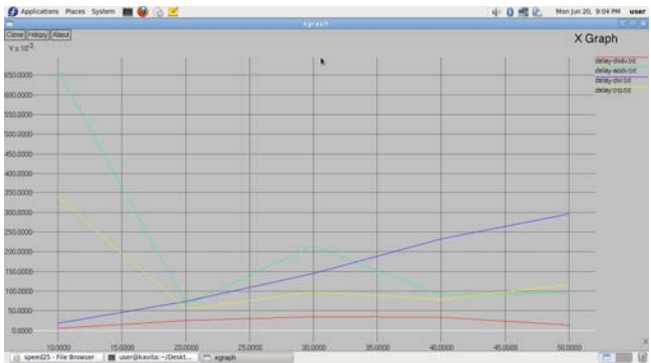
Graph 9.2: Average Delay, Speed 20 and varying number of nodes

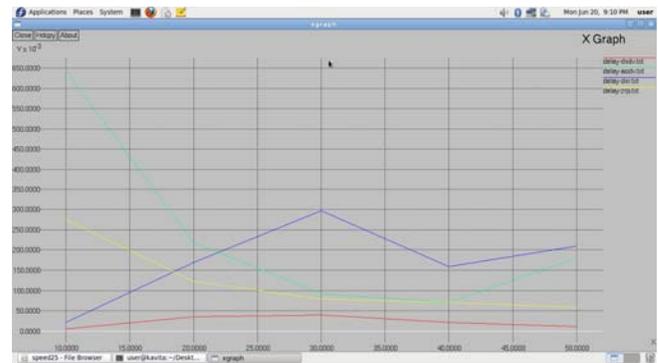
Graph 10.2: Average Delay, Speed 25 and varying number of nodes

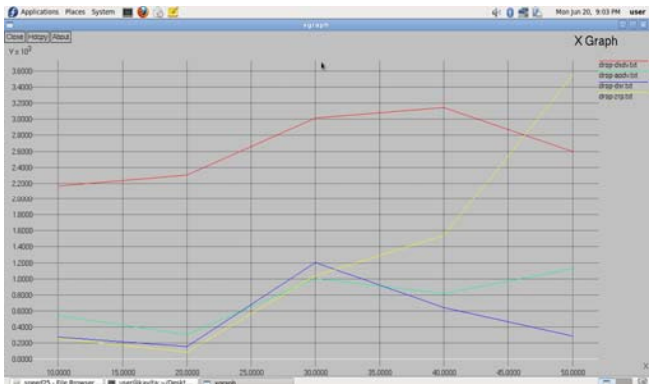
Graph 9.3: Dropped Packets, Speed 20 and varying number of nodes

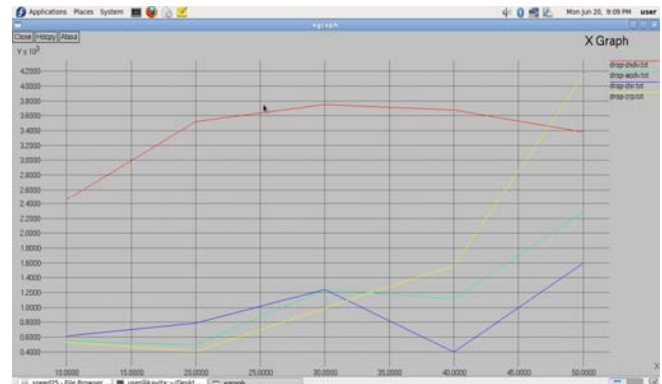
Graph 10.3: Dropped Packets, Speed 25 and varying number of nodes

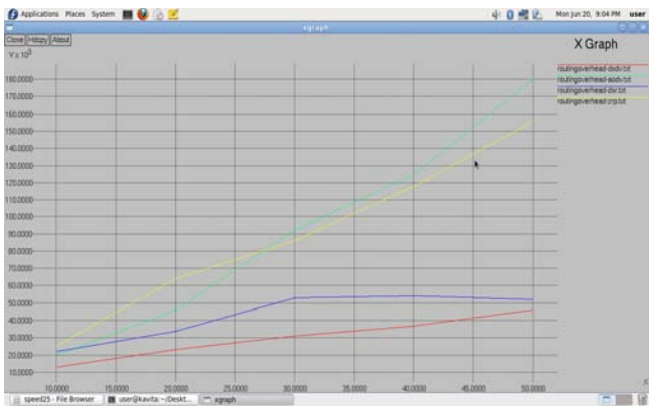
Graph 9.4: Routing Overhead, Speed 20 and varying number of nodes

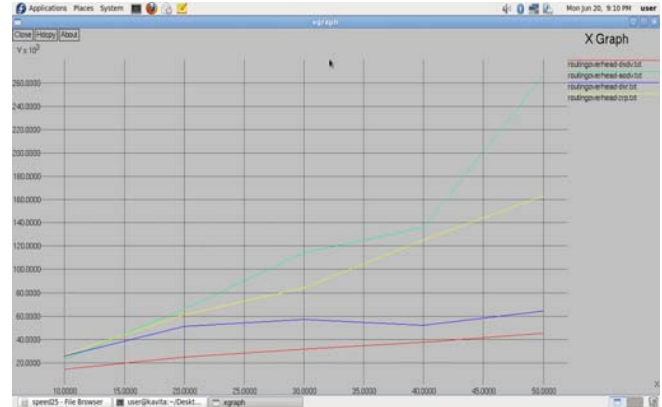
Graph 10.4: Routing Overhead, Speed 25 and varying number of nodes

## 5. Conclusion

In the present exposition, the performance of MANET routing protocols is examined with respect to following four performance metrics namely, throughput, average delay, number of packets dropped and routing overhead. DSDV is a proactive protocol, where as, AODV and DSR falls under the category of reactive protocol and ZRP is a hybrid protocol. The simulation results suggest that each protocol performs well in some scenario yet has some drawbacks in other cases. In terms of throughput, AODV performance is better than others whereas, DSDV performance poorly sometimes. Another disadvantage of DSDV is that the number of dropped packets is also significantly higher. ZRP throughput does not change even with a change in mobility or pause time because of its hybrid nature. The performance of DSR is good in terms of routing overhead and number of packets dropped due to route cache. It can also be concluded from the simulation results that the reliability of AODV and DSR protocols is better than other two protocols.

## 6. References

[1]. Kapang Lego, Pranav Kumar Singh, Dipankar Sutradhar, *"Comparative Study of Adhoc Routing Protocol AODV, DSR and DSDV in Mobile Adhoc NETwork"*, Indian Journal of Computer Science and Engineering Vol. 1 No. 4 364-371, 2011.
[2]. S.R. Birdar, Hiren H D Sarma, Kalpana Sharma, Subir Kumar Sarkar, Puttamadappa C, *Performance Comparison of Reactive Routing Protocols of MANETs using Group Mobility Model"*, International Conference on Signal Processing Systems, 2009.
[3]. G. Jayakumar and G. Gopinath, *"Performance comparison of two on-demand routing protocols for ad-hoc networks based on random way point mobility model,"* American Journal of Applied Sciences, vol. 5, no. 6, pp. 649-664, June 2008.
[4]. S. Ahmed and M. S. Alam, *"Performance Evaluation of important ad hoc networks protocols"*, EURASIP Journal on wireless Communications and networking, Vol: 2006, Article ID 78645, PP 1-11, 2006.
[5]. Guntupalli Lakshmikant, A Gaiwak, P.D. Vyavahare, *"Simulation Based Comparative Performance Analysis of Adhoc Routing Protocols"*, in proceedings of TENCON 2008.
[6]. OLSR, internet draft, http://tools.ietf.org/html/draft-ietf-manet-olsr-00
[7]. G. Vijaya Kumar, Y. Vasudeva Reddyr, M. Nagendra, *"Current Research Work on Routing Protocols for MANET: A Literature Survey"*, International Journal on Computer Science and Engineering, Vol. 02, No. 03, pp. 706-713, 2010.
[8]. Vincent D. Park, M. Scott Corson, Temporally-Ordered Routing Algorithm (TORA) version 1: functional specification, Internet-Draft, draft-ietf-manet-tora-spec- 00.txt, November 1997.
[9]. V. Ramasubramanian, Z. J. Haas, and E. G. Sirer, "SHARP: A Hybrid Adaptive Routing Protocol for Mobile Ad Hoc Networks," The Fourth ACM International Symposium on Mobile Ad Hoc Networking and Computing (MobiHoc), pp. 303-314, 2003.
[10]. Yogesh Chaba, Yudhvir Singh, Manish Joon, *"Simulation Based Performance Analysis of On-Demand Routing Protocols in MANETs,"* Second International Conference on Computer Modeling and Simulation, 2010.
[11]. Chenna Reddy, P.; ChandraSekhar Reddy, P., *"Performance Analysis of Adhoc Network Routing Protocols,"* ISAUHC '06, International Symposium on Ad Hoc and Ubiquitous Computing, vol., no., pp.186-187, 20-23 Dec. 2006.
[12]. Vijayalaskhmi M. Avinash Patel, Linganagouda Kulkarni, *"QoS Parameter Analysis on AODV and DSDV Protocols in a Wireless Network"*, International Journal of Communication Network and Security, Volume-1, Issue-1, 2011.
[13]. Shaily Mittal, Prabhjot Kaur, *"Performance Comparison of AODV, DSR and ZRP Routing Protocols in MANET's"*, International Conference on Advances in Computing, Control, and Telecommunication Technologies, 2009.
[14]. Li Layuan, Li Chunlin, Yaun Peiyan, "Performance evaluation and simulation of routing protocols in ad hoc networks", Computer Communications 30 (2007) 1890-1898.
[15]. C.K. Toh, Ad Hoc Mobile Wireless Networks Protocols and Systems, Pearson Education, 2007.
[16]. Sunil Taneja, Ashwani Kush, *"A Survey of Routing Protocols in Mobile Adhoc Networks"*, International Journal of Innovation, Management and Technology, Vol. 1, No. 3, August 2010.
[17]. Charles E. Perkins and Pravin Bhagwat, "Highly Dynamic Destination-Sequenced Distance-Vector routing (dsdv) for Mobile Computers", 1994.
[18]. DSR, internet draft, http://tools.ietf.org/html/draft-ietf-manet-dsr-10 .
[19]. AODV, internet draft, http://tools.ietf.org/html/draft-ietf-manet-aodv-09 .
[20]. ZRP, internet draft, http://tools.ietf.org/id/draft-ietf-manet-zone-zrp-04.txt .
[21]. ZRP patch, http://magnet.daiict.ac.in/magnet_members/MTech/2007/PatelBrijesh/Simulation.html#Sec_2.
[22]. ZRP Agent Implementation documentation, http://magnet.daiict.ac.in/magnet_members/MTech/2007/PatelBrijesh/Thesis_files/MyZRP/ZRPManual.pdf .
[23]. Yinfei Pan, *"Design Routing Protocol Performance Comparison in NS2: AODV Comparing to DSR as Example"*, Deptt of CS, SUNY Binghamton, Vestal NY 13850.
[24]. NS2 Trace format - http://nsnam.isi.edu/nsnam/index.php/NS-2_Trace_Formats .
[25]. The ns Manual (formerly ns Notes and Documentation) by Kevin Fall, Kannan Varadhan. http://www.isi.edu/nsnam/ns/doc/ns_doc.pdf
[26]. NS Simulator for beginners, http://www-sop.inria.fr/members/Eitan.Altman/COURS-NS/n3.pdf.

**Kavita Pandey** She did B.Tech in Computer Science from M.D. University in year 2002 and M.Tech. in Computer Science from Banasthali Vidyapith University in year 2003. She is pursuing PhD from JIIT,Noida. She is working as a Senior Lecturer in JIIT,Noida. Her current research interests include Adhoc Networks, Optimization Techniques and Network Security.

**Abhishek Swaroop** He received his B.Tech in computer science and engineering from G.B.Pant Univ. Pantnagar in 1992, M.Tech from Punajbi Univ. Patiala in 2004 and Ph.D. in computer engineering from N.I.T. Kurukshetra in 2011. He is currently working as an Assitant professor in JIIT, Noida. His research interests include group mutual exclusion, fault tolerance, MANETs, Sensor networks, Multi core architecture.